\documentclass[twocolumn,letterpaper,aps,prl,longbibliography,superscriptaddress,showpacs,floatfix]{revtex4-1}

\usepackage{graphicx}	
\usepackage{amsmath}
\usepackage{xspace}	
\usepackage{color}
\usepackage[mathscr]{euscript}

\newcommand{\s}[1]{\mbox{$\sqrt{s}$ = #1\,GeV}}

\newcommand{\pp}{$p$+$p$\ }

\newcommand{\pt}{\mbox{$p_T$}\xspace}

\newcommand{\ee}{\mbox{$ee$}\xspace}
\newcommand{\emu}{\mbox{$e\mu$}\xspace}

\newcommand{\fraction}{\mbox{$\textbf{F}$}\xspace}
\newcommand{\n}{\mbox{$\textbf{n}$}\xspace}

\newcommand{\D}{\mbox{$\textbf{D}$}\xspace}
\newcommand{\sigmahf}{\mbox{$\sigma_{\rm HF}$}\xspace}

\newcommand{\mumu}{\mbox{$\mu\mu$}\xspace}
\newcommand{\lsmumu}{\mbox{$\mu^{\pm}\mu^{\pm}$}\xspace}
\newcommand{\ulmumu}{\mbox{$\mu^{+}\mu^{-}$}\xspace}

\newcommand{\cc}{$c\bar{c}$\xspace}
\newcommand{\bb}{$b\bar{b}$\xspace}

\newcommand{\powheg}{{\sc powheg}\xspace}
\newcommand{\pythia}{{\sc pythia}\xspace}

\begin{document}

\title{Correlations of $\mu\mu$, $e\mu$, and $ee$ pairs in $p$+$p$ 
collisions at $\sqrt{s}=200$ GeV and implications for $c\bar{c}$ and 
$b\bar{b}$ production mechanisms}

\newcommand{\abilene}{Abilene Christian University, Abilene, Texas 79699, USA}
\newcommand{\augie}{Department of Physics, Augustana University, Sioux Falls, South Dakota 57197, USA}
\newcommand{\banaras}{Department of Physics, Banaras Hindu University, Varanasi 221005, India}
\newcommand{\barc}{Bhabha Atomic Research Centre, Bombay 400 085, India}
\newcommand{\baruch}{Baruch College, City University of New York, New York, New York, 10010 USA}
\newcommand{\bnlcoll}{Collider-Accelerator Department, Brookhaven National Laboratory, Upton, New York 11973-5000, USA}
\newcommand{\bnlphys}{Physics Department, Brookhaven National Laboratory, Upton, New York 11973-5000, USA}
\newcommand{\caucr}{University of California-Riverside, Riverside, California 92521, USA}
\newcommand{\charlesczech}{Charles University, Ovocn\'{y} trh 5, Praha 1, 116 36, Prague, Czech Republic}
\newcommand{\chonbuk}{Chonbuk National University, Jeonju, 561-756, Korea}
\newcommand{\cns}{Center for Nuclear Study, Graduate School of Science, University of Tokyo, 7-3-1 Hongo, Bunkyo, Tokyo 113-0033, Japan}
\newcommand{\colorado}{University of Colorado, Boulder, Colorado 80309, USA}
\newcommand{\columbia}{Columbia University, New York, New York 10027 and Nevis Laboratories, Irvington, New York 10533, USA}
\newcommand{\czechtech}{Czech Technical University, Zikova 4, 166 36 Prague 6, Czech Republic}
\newcommand{\debrecen}{Debrecen University, H-4010 Debrecen, Egyetem t{\'e}r 1, Hungary}
\newcommand{\elte}{ELTE, E{\"o}tv{\"o}s Lor{\'a}nd University, H-1117 Budapest, P{\'a}zm{\'a}ny P.~s.~1/A, Hungary}
\newcommand{\eszterhazy}{Eszterh\'azy K\'aroly University, K\'aroly R\'obert Campus, H-3200 Gy\"ongy\"os, M\'atrai \'ut 36, Hungary}
\newcommand{\ewha}{Ewha Womans University, Seoul 120-750, Korea}
\newcommand{\fsu}{Florida State University, Tallahassee, Florida 32306, USA}
\newcommand{\gsu}{Georgia State University, Atlanta, Georgia 30303, USA}
\newcommand{\hiroshima}{Hiroshima University, Kagamiyama, Higashi-Hiroshima 739-8526, Japan}
\newcommand{\howard}{Department of Physics and Astronomy, Howard University, Washington, DC 20059, USA}
\newcommand{\ihepprot}{IHEP Protvino, State Research Center of Russian Federation, Institute for High Energy Physics, Protvino, 142281, Russia}
\newcommand{\illuiuc}{University of Illinois at Urbana-Champaign, Urbana, Illinois 61801, USA}
\newcommand{\inrras}{Institute for Nuclear Research of the Russian Academy of Sciences, prospekt 60-letiya Oktyabrya 7a, Moscow 117312, Russia}
\newcommand{\instpasczech}{Institute of Physics, Academy of Sciences of the Czech Republic, Na Slovance 2, 182 21 Prague 8, Czech Republic}
\newcommand{\isu}{Iowa State University, Ames, Iowa 50011, USA}
\newcommand{\jaea}{Advanced Science Research Center, Japan Atomic Energy Agency, 2-4 Shirakata Shirane, Tokai-mura, Naka-gun, Ibaraki-ken 319-1195, Japan}
\newcommand{\kek}{KEK, High Energy Accelerator Research Organization, Tsukuba, Ibaraki 305-0801, Japan}
\newcommand{\korea}{Korea University, Seoul, 136-701, Korea}
\newcommand{\kurchatov}{National Research Center ``Kurchatov Institute", Moscow, 123098 Russia}
\newcommand{\kyoto}{Kyoto University, Kyoto 606-8502, Japan}
\newcommand{\lawllnl}{Lawrence Livermore National Laboratory, Livermore, California 94550, USA}
\newcommand{\losalamos}{Los Alamos National Laboratory, Los Alamos, New Mexico 87545, USA}
\newcommand{\lund}{Department of Physics, Lund University, Box 118, SE-221 00 Lund, Sweden}
\newcommand{\lyon}{IPNL, CNRS/IN2P3, Univ Lyon, Université Lyon 1, F-69622, Villeurbanne, France}
\newcommand{\maryland}{University of Maryland, College Park, Maryland 20742, USA}
\newcommand{\mass}{Department of Physics, University of Massachusetts, Amherst, Massachusetts 01003-9337, USA}
\newcommand{\michigan}{Department of Physics, University of Michigan, Ann Arbor, Michigan 48109-1040, USA}
\newcommand{\muhlenberg}{Muhlenberg College, Allentown, Pennsylvania 18104-5586, USA}
\newcommand{\nara}{Nara Women's University, Kita-uoya Nishi-machi Nara 630-8506, Japan}
\newcommand{\natmephi}{National Research Nuclear University, MEPhI, Moscow Engineering Physics Institute, Moscow, 115409, Russia}
\newcommand{\newmex}{University of New Mexico, Albuquerque, New Mexico 87131, USA}
\newcommand{\nmsu}{New Mexico State University, Las Cruces, New Mexico 88003, USA}
\newcommand{\ohio}{Department of Physics and Astronomy, Ohio University, Athens, Ohio 45701, USA}
\newcommand{\ornl}{Oak Ridge National Laboratory, Oak Ridge, Tennessee 37831, USA}
\newcommand{\orsay}{IPN-Orsay, Univ.~Paris-Sud, CNRS/IN2P3, Universit\'e Paris-Saclay, BP1, F-91406, Orsay, France}
\newcommand{\peking}{Peking University, Beijing 100871, People's Republic of China}
\newcommand{\pnpi}{PNPI, Petersburg Nuclear Physics Institute, Gatchina, Leningrad region, 188300, Russia}
\newcommand{\riken}{RIKEN Nishina Center for Accelerator-Based Science, Wako, Saitama 351-0198, Japan}
\newcommand{\rikjrbrc}{RIKEN BNL Research Center, Brookhaven National Laboratory, Upton, New York 11973-5000, USA}
\newcommand{\rikkyo}{Physics Department, Rikkyo University, 3-34-1 Nishi-Ikebukuro, Toshima, Tokyo 171-8501, Japan}
\newcommand{\saispbstu}{Saint Petersburg State Polytechnic University, St.~Petersburg, 195251 Russia}
\newcommand{\seoulnat}{Department of Physics and Astronomy, Seoul National University, Seoul 151-742, Korea}
\newcommand{\stonybrkc}{Chemistry Department, Stony Brook University, SUNY, Stony Brook, New York 11794-3400, USA}
\newcommand{\stonycrkp}{Department of Physics and Astronomy, Stony Brook University, SUNY, Stony Brook, New York 11794-3800, USA}
\newcommand{\tenn}{University of Tennessee, Knoxville, Tennessee 37996, USA}
\newcommand{\titech}{Department of Physics, Tokyo Institute of Technology, Oh-okayama, Meguro, Tokyo 152-8551, Japan}
\newcommand{\tsukuba}{Tomonaga Center for the History of the Universe, University of Tsukuba, Tsukuba, Ibaraki 305, Japan}
\newcommand{\vandy}{Vanderbilt University, Nashville, Tennessee 37235, USA}
\newcommand{\weizmann}{Weizmann Institute, Rehovot 76100, Israel}
\newcommand{\wigner}{Institute for Particle and Nuclear Physics, Wigner Research Centre for Physics, Hungarian Academy of Sciences (Wigner RCP, RMKI) H-1525 Budapest 114, POBox 49, Budapest, Hungary}
\newcommand{\yonsei}{Yonsei University, IPAP, Seoul 120-749, Korea}
\newcommand{\zagreb}{Department of Physics, Faculty of Science, University of Zagreb, Bijeni\v{c}ka c.~32 HR-10002 Zagreb, Croatia}
\affiliation{\abilene}
\affiliation{\augie}
\affiliation{\banaras}
\affiliation{\barc}
\affiliation{\baruch}
\affiliation{\bnlcoll}
\affiliation{\bnlphys}
\affiliation{\caucr}
\affiliation{\charlesczech}
\affiliation{\chonbuk}
\affiliation{\cns}
\affiliation{\colorado}
\affiliation{\columbia}
\affiliation{\czechtech}
\affiliation{\debrecen}
\affiliation{\elte}
\affiliation{\eszterhazy}
\affiliation{\ewha}
\affiliation{\fsu}
\affiliation{\gsu}
\affiliation{\hiroshima}
\affiliation{\howard}
\affiliation{\ihepprot}
\affiliation{\illuiuc}
\affiliation{\inrras}
\affiliation{\instpasczech}
\affiliation{\isu}
\affiliation{\jaea}
\affiliation{\kek}
\affiliation{\korea}
\affiliation{\kurchatov}
\affiliation{\kyoto}
\affiliation{\lawllnl}
\affiliation{\losalamos}
\affiliation{\lund}
\affiliation{\lyon}
\affiliation{\maryland}
\affiliation{\mass}
\affiliation{\michigan}
\affiliation{\muhlenberg}
\affiliation{\nara}
\affiliation{\natmephi}
\affiliation{\newmex}
\affiliation{\nmsu}
\affiliation{\ohio}
\affiliation{\ornl}
\affiliation{\orsay}
\affiliation{\peking}
\affiliation{\pnpi}
\affiliation{\riken}
\affiliation{\rikjrbrc}
\affiliation{\rikkyo}
\affiliation{\saispbstu}
\affiliation{\seoulnat}
\affiliation{\stonybrkc}
\affiliation{\stonycrkp}
\affiliation{\tenn}
\affiliation{\titech}
\affiliation{\tsukuba}
\affiliation{\vandy}
\affiliation{\weizmann}
\affiliation{\wigner}
\affiliation{\yonsei}
\affiliation{\zagreb}
\author{C.~Aidala} \affiliation{\michigan} 
\author{Y.~Akiba} \email[PHENIX Spokesperson: ]{akiba@rcf.rhic.bnl.gov} \affiliation{\riken} \affiliation{\rikjrbrc} 
\author{M.~Alfred} \affiliation{\howard} 
\author{V.~Andrieux} \affiliation{\michigan} 
\author{N.~Apadula} \affiliation{\isu} 
\author{H.~Asano} \affiliation{\kyoto} \affiliation{\riken} 
\author{B.~Azmoun} \affiliation{\bnlphys} 
\author{V.~Babintsev} \affiliation{\ihepprot} 
\author{A.~Bagoly} \affiliation{\elte} 
\author{N.S.~Bandara} \affiliation{\mass} 
% ...
\collaboration{PHENIX Collaboration}  \noaffiliation

\date{\today}


\begin{abstract}

PHENIX has measured the azimuthal correlations of muon pairs from charm 
and bottom semi-leptonic decays in $p$+$p$ collisions at $\sqrt{s}=200$ 
GeV, using a novel analysis technique utilizing both unlike- and 
like-sign muon pairs to separate charm, bottom and Drell-Yan 
contributions. The dimuon measurements combined with the previous 
electron-muon and dielectron measurements span a wide range in rapidity, 
and are well described by \pythia Tune A. Through a Bayesian analysis 
based on \pythia Tune A, we show that leading order pair creation is the 
dominant $(76\%\pm^{14}_{19}\%)$ contribution for \bb production, 
whereas the data favor the scenario in which next-to-leading-order 
processes dominate \cc production. The small contribution of 
next-to-leading-order processes in \bb production at the collision 
energies of the Relativistic Heavy Ion Collider contrasts with the case 
at Large-Hadron-Collider energies, where next-to-leading-order processes 
are expected to dominate.
 
\end{abstract}

\maketitle




Despite substantial experimental and theoretical efforts in recent 
years, our understanding of heavy flavor production in \pp collisions 
remains incomplete. Differential cross section measurements, 
particularly for charm, are systematically higher than the central 
values of theoretical 
predictions~\cite{Cacciari:1998it,Sjostrand:2006za,Norrbin:2000zc,Frixione:2007nw,Frixione:2003ei,Vogt:2007aw} 
for collision energies from the Relativistic Heavy Ion
Collider (RHIC)~\cite{Adare:2010de,Aidala:2017pum,Xie:2017nal} 
to the Large Hadron Collider 
(LHC)~\cite{Acosta:2003ax,Acharya:2017jgo,Aaij:2015bpa,Aad:2015zix,Sirunyan:2017xss}, 
and are only consistent when large theoretical uncertainties are 
considered.

Angular correlations of quarks and anti-quarks are a unique probe for 
studying heavy flavor production in \pp collisions. Leading-order (LO) 
pair-creation processes feature a strong back-to-back azimuthal angular 
correlation, while the distributions from next-to-leading order (NLO) 
processes are broader~\cite{Norrbin:2000zc,Ilten:2017rbd}. Thus, 
relative contributions from different production mechanisms can be 
disentangled by studying the azimuthal angular correlations of heavy 
mesons or their decay products. As the fraction of NLO processes is 
expected to increase with beam energy~\cite{Norrbin:2000zc}, angular 
correlations provide an important handle for investigating the energy 
dependence of heavy flavor production.

Only a few heavy-flavor correlation measurements have been performed at 
high energies. At the Tevatron~\cite{Acosta:2004nj} and the 
LHC~\cite{Khachatryan:2011wq,Aaij:2012dz} data are reasonably well 
described by NLO perturbative quantum chromodynamics (pQCD) 
calculations, but only a few quantitative constraints have been extracted 
on the relative contributions of different heavy-flavor production 
mechanisms.  At RHIC, inclusive heavy flavor (dominated by \cc) 
\ee~\cite{ppg189} and \emu~\cite{ppg130} measurements at 
mid-midrapidity and mid-forward rapidity in \pp collisions at \s{200} are 
consistent with pQCD models within experimental uncertainties. However, 
the limited statistical accuracy of these measurements prohibit us from 
providing strong constraints on heavy flavor production mechanisms.

PHENIX~\cite{Adcox:2003zm} has recently measured azimuthal correlations 
of \mumu pairs from \cc and \bb~\cite{ppg213} in \pp collisions at 
$\sqrt{s}=200$ GeV, using the high statistics data set taken in 2015 
that corresponds to an integrated luminosity of $\int \mathcal{L}dt=51 $ 
pb$^{-1}$. The \mumu data, together with the previous \ee and \emu 
measurements cover a wide kinematic range. Here, we present an analysis 
of \cc and \bb correlations in $p$+$p$ collisions at \s{200}, where we 
combine the \mumu, \emu and \ee measurements to constrain the \cc and 
\bb production mechanisms.

\begin{figure}[htb]
\includegraphics[width=1.0\linewidth]{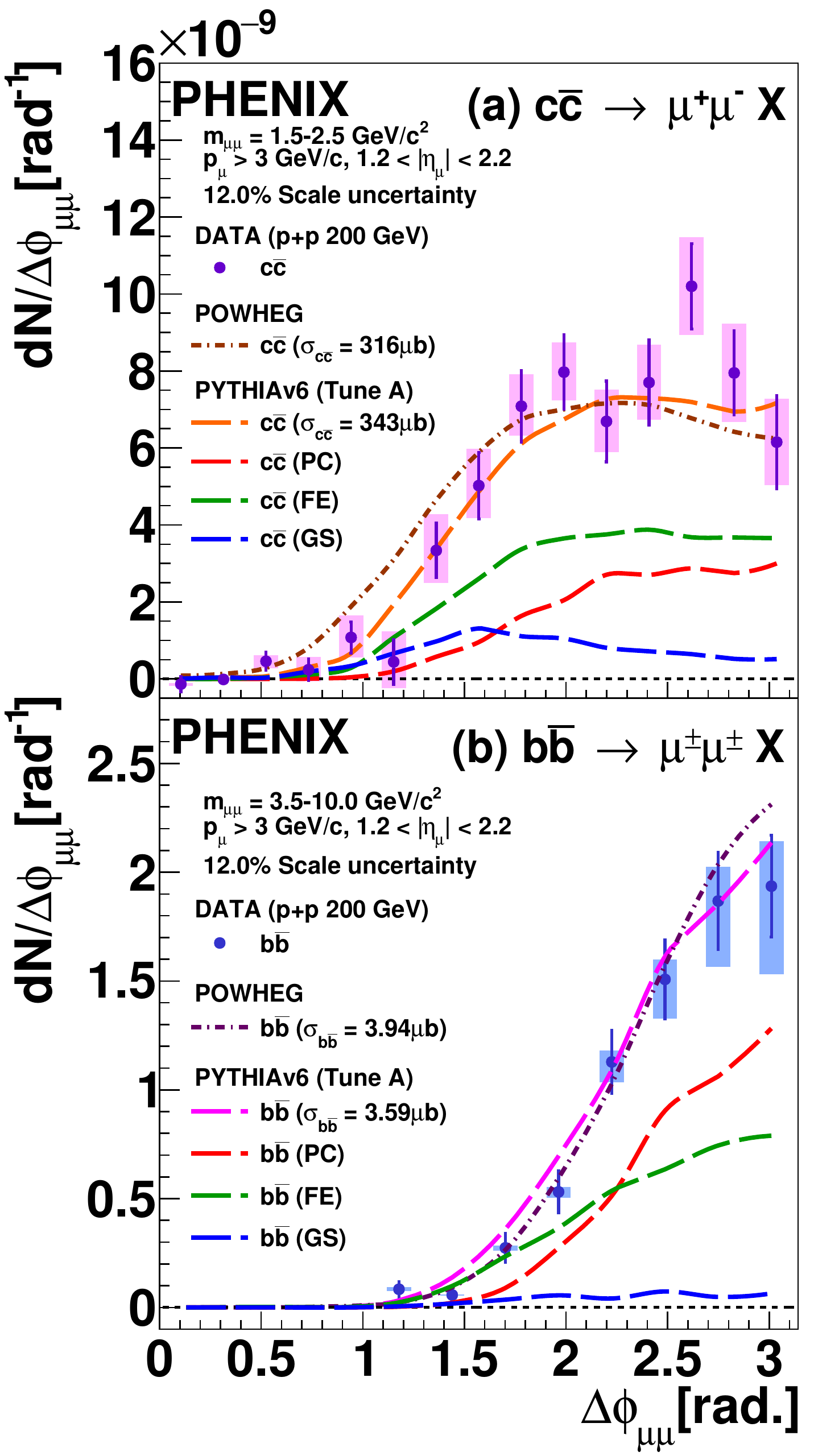}
\caption{\label{fig:hfleptonsmmccbb} 
Dimuon azimuthal correlations from \cc(a) and \bb(b) compared to \pythia 
and \powheg. }
\end{figure}

A complete description of the \mumu analysis can be found 
in~\cite{ppg213}. The \mumu pairs from \cc, \bb, Drell-Yan and hadronic 
pairs (arising from kaons and pions) are separated via a simultaneous 
fit to unlike- and like-sign pairs in mass and transverse momentum \pt. 
The highlight of the analysis is the extraction of azimuthal 
correlations of \mumu from \bb utilizing like-sign pairs. Decays from 
\cc or the Drell-Yan mechanism result in unlike-sign pairs only; in 
contrast \bb can result in like-sign pairs either via a combination of 
$B \rightarrow \mu$ and $B\rightarrow D\rightarrow \mu$ decay chains or 
decays following $B^{0}\bar{B^{0}}$ oscillations. These pairs dominate 
the high mass like-sign spectrum ($3.5<m_{\mu\mu}$[GeV$/c^{2}]<10.0$), 
which allows isolating a sample of dimuons from \bb and hadronic pairs 
with S/B $\sim1$.  The hadronic pairs are subtracted, and the remaining 
\mumu pairs are corrected for efficiency to obtain the \bb yields.

Figure~\ref{fig:hfleptonsmmccbb} shows the \mumu pair yield from \cc and 
\bb separately as a function of the azimuthal angle between the two 
muons. Distributions from the event generators, \pythia 
v6.428~\cite{Sjostrand:2006za} and \powheg v1.0~\cite{Frixione:2007nw} 
(interfaced with \pythia v8.100~\cite{Sjostrand:2007gs}) are compared to 
the data. For \pythia, contributions from pair creation (PC), flavor 
excitation (FE), and gluon splitting (GS)~\cite{Norrbin:2000zc} are 
shown separately. \pythia and \powheg treat the NLO corrections 
differently: \pythia implements NLO corrections with a parton-shower 
approach, while NLO corrections are directly implemented in the hard 
process using NLO matrix elements in \powheg. Tune A 
parameters~\cite{Field:2002vt} are used for \pythia; default settings 
are used for \powheg. Details on the simulation setup can be found 
in~\cite{ppg213}. Generated distributions are normalized using cross 
sections obtained in the fitting procedure documented in~\cite{ppg213} 
($\sigma_{c\bar{c}}=343~\mu$b, $\sigma_{b\bar{b}}=3.59~\mu$b for 
\pythia, $\sigma_{c\bar{c}}=316~\mu$b, $\sigma_{b\bar{b}}=3.94~\mu$b for 
\powheg).

\begin{figure}[htb]
\includegraphics[width=1.0\linewidth]{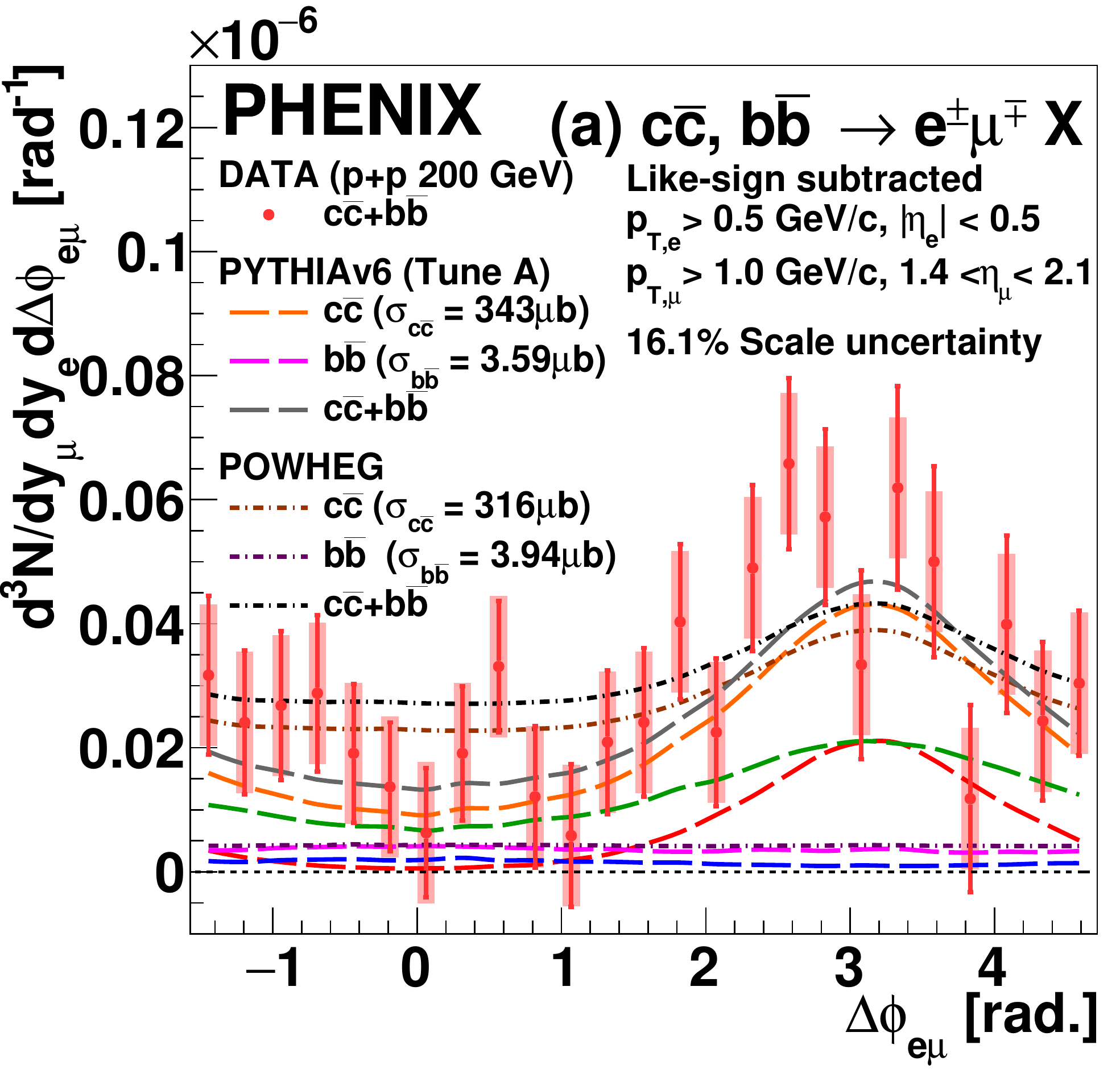}
\includegraphics[width=1.0\linewidth]{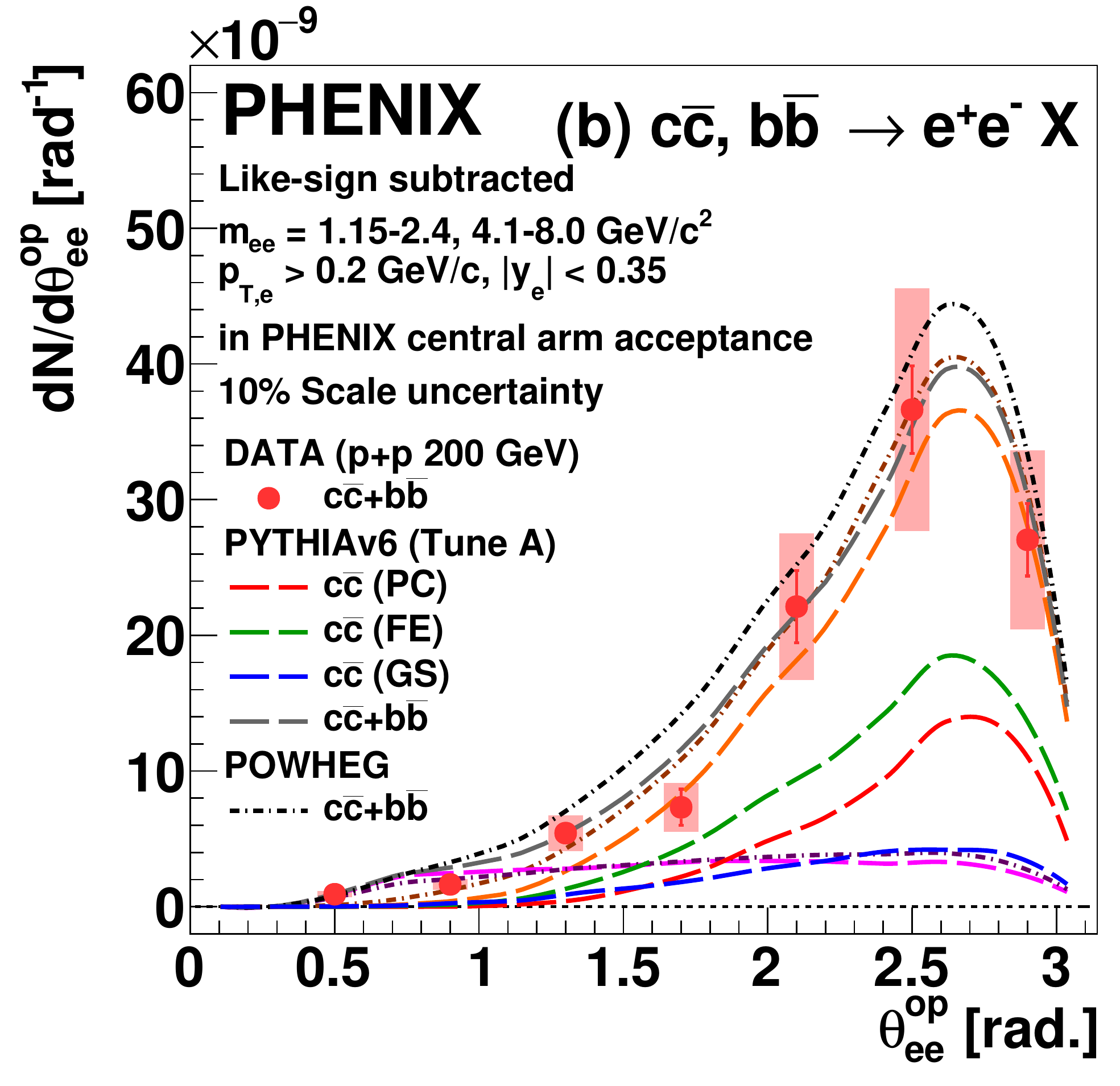}
\caption{\label{fig:hfleptonsemeeccbb} 
Electron-muon azimuthal correlations from \cc and \bb(a), and dielectron 
opening angle distributions from \cc and \bb(b), compared to \pythia and 
\powheg. The legends are shared among panels (a) and (b).} 
\end{figure}

Figure~\ref{fig:hfleptonsemeeccbb} shows the measured heavy flavor 
\ee~\cite{ppg189} and \emu~\cite{ppg130} yields, as a function of the 
azimuthal opening angle $\Delta\phi_{e\mu}$ and the opening angle 
$\theta_{ee}^{op}$, respectively. These yields are extracted in 
distinctly different (pseudo)rapidity regions ($e\mu:|\eta_{e}|<0.5, 
1.4<|\eta_{\mu}|<2.1$; $ee:|y_{e}|<0.35$) compared to the \mumu pairs 
($\mu\mu:1.2<|\eta_{\mu}|<2.2$). The \emu pairs contain contributions 
from \cc and \bb, while the \ee pairs contain also additional but 
negligible ($<0.5\%$) contributions from Drell-Yan. Distributions from 
\cc and \bb generated from \pythia and \powheg are normalized using the 
cross sections obtained in the \mumu analysis ~\cite{ppg213}, and 
compared to data. In both cases, the yield is dominated by pairs from 
\cc.

Although the correlations of the lepton pairs are measured within 
limited detector acceptance and have additional kinematic constraints, a 
strong back-to-back peak is observed for leading order PC for both \cc 
and \bb. Distributions from FE and GS are significantly broader than 
those from PC. To quantify the consistency with data, we calculate a 
modified $\tilde{\chi}^{2}$~\cite{Adare:2008cg} that takes systematic 
uncertainties into account. For \bb, the $\tilde{\chi}^{2}/NDF$ values 
for \pythia and \powheg are $9.8/7$ and $7.2/7$, respectively, which 
indicates that the the azimuthal correlations for \bb are well described 
by both models. For \cc, the $\tilde{\chi}^{2}/NDF$ values of \pythia 
and \powheg are $20.1/14$ and $35.8/14$, respectively. While the \mumu 
data are well described by \pythia, the distribution from \powheg are 
wider than in the data. The $\tilde{\chi}^{2}/NDF$ value obtained by 
comparing \pythia to the \cc dominated \ee and \emu measurements and the 
\cc only \mumu measurement is $59.6/47$. This indicates \pythia can 
describe both the rapidity dependence and angular correlations of \cc 
production well. The corresponding $\tilde{\chi}^{2}/NDF$ value for 
\powheg is $94.2/47$. 

Because distributions of decay lepton pairs are highly correlated to the 
$c$ and $\bar{c}$ quarks~\cite{ppg189}, this indicates that the 
description of $c$-$\bar{c}$ quark correlations between \pythia and 
\powheg is intrinsically different at the quark level. In addition, we 
observe that at $\Delta\phi<\pi/2$ which is dominated by NLO processes, 
\powheg always predicts more yield than \pythia; while the ratio of the 
yields at $\Delta\phi>\pi/2$ of \powheg to \pythia decreases with 
rapidity in the measured phase spaces.  Because leading order processes are 
peaked near $\Delta\phi=\pi$, this may imply that the rapidity 
dependence of the ratio of LO to NLO contributions is different between 
the two models. 

To further constrain the production mechanisms of \cc and \bb, we 
perform a simultaneous shape analysis of the \mumu, \emu, and \ee data 
shown in Figs.~\ref{fig:hfleptonsmmccbb} and \ref{fig:hfleptonsemeeccbb} 
using Bayesian inference. Because the measurements cover different parts 
of phase space, extrapolations are unavoidable. \pythia Tune A gives 
good agreement with multiple measurements made at the 
Tevatron~\cite{Field:2008zz}, as well as jet and underlying event 
measurements from PHENIX~\cite{Adare:2010cc} and 
STAR~\cite{Adamczyk:2012qj}; we thus focus on Tune A for this study.

The analysis is performed separately for \cc and \bb. For \bb, we only 
use the \mumu data set, whereas for \cc, the \ee, \emu and \mumu data 
sets are used. For \ee and \emu data, we first subtract the expected \bb 
yield from the two data sets and assign additional systematic 
uncertainties on extrapolation $(\sim2\%)$ and normalization 
$(\sim6\%)$. The extrapolation uncertainties are estimated by taking the 
difference between \pythia and \powheg; the uncertainties on the 
normalization are taken from~\cite{ppg213}.

For \cc or \bb, the model prediction of the yield, 
$\textbf{T}=\{T_{i,j}\}$ for the $i^{th}$ data set, either \ulmumu, 
\emu, or \ee for \cc and \lsmumu for \bb, in the $j^{th}$ (azimuthal) 
opening angle bin can be written as:

\begin{align}
\label{eq:factorization}
T_{i,j}(\fraction,\sigmahf) = \sigmahf \sum_{\alpha} f_{\alpha}Y_{\alpha,i,j},
\end{align}

\noindent where $\fraction=(F_{\rm PC},F_{\rm FE},F_{\rm GS})$ is the relative 
contribution to heavy flavor production in $4\pi$ phase space from the 
three considered processes PC, FE, and GS, \sigmahf is the total heavy 
flavor cross section in $4\pi$, and $Y_{\alpha,i,j}$ is the yield in the 
measured phase space of the $i^{th}$ data set (indicated in 
Figs.~\ref{fig:hfleptonsmmccbb} and~\ref{fig:hfleptonsemeeccbb}) for the 
$j^{th}$ bin per event generated involving the $\alpha$ process, where 
$\alpha=$ PC, FE or GS. The quantity that we constrain from the data is 
the relative contribution \fraction, which is directly related to the 
shape of the angular distributions. The total heavy flavor cross section 
\sigmahf sets the overall normalization and is unimportant for this 
shape analysis.

The shape analysis of the angular distributions is sensitive to 
systematic uncertainties. The background subtraction is the dominant 
source of systematic uncertainty for all lepton-pair combinations. It 
introduces systematic uncertainties of ${\sim}20$\% for \mumu from 
\cc~\cite{ppg213}, ${\sim}15$\% for \mumu from \bb~\cite{ppg213}, 
${\sim}30$\% for \emu~\cite{ppg130}, and ${\sim}20$\% for \ee~\cite{ppg189}, 
which affects the data points in a correlated manner.  We adopt a 
Bayesian approach to account for these systematic variations,

Based on Eq.~\ref{eq:factorization}, we can construct a vector of model 
parameters {\protect\boldmath$\theta$}, which comprise the relative 
fractions of heavy flavor production processes \fraction and the heavy 
flavor cross section \sigmahf.  In the Bayesian approach, systematic 
uncertainties are naturally accounted for by incorporating nuisance 
parameters \n into {\protect\boldmath$\theta$}, where each nuisance 
parameter corresponds to one source of systematic uncertainty 
(see~\cite{berger1999} for a pedagogical review).  From Bayes' rule, one 
can write:

\begin{align}
\label{eq:bayes1}
P(\fraction,\sigmahf,\n|D)=\frac{P(\D|\fraction,\sigmahf,\n)\cdot P(\fraction,\sigmahf,\n)}{P(\D)}.
\end{align}

\noindent The quantity that we want to obtain is $P(\fraction|\D)$. We 
assume a noninformative prior for \fraction, i.e. a uniform distribution 
in the \textit{physical region}, in which the values $F_{i}$, where 
$i=$PC, FE, GS, lie between zero and one and sum to one.

\begin{figure*}[thb]
\includegraphics[width=0.329\linewidth]{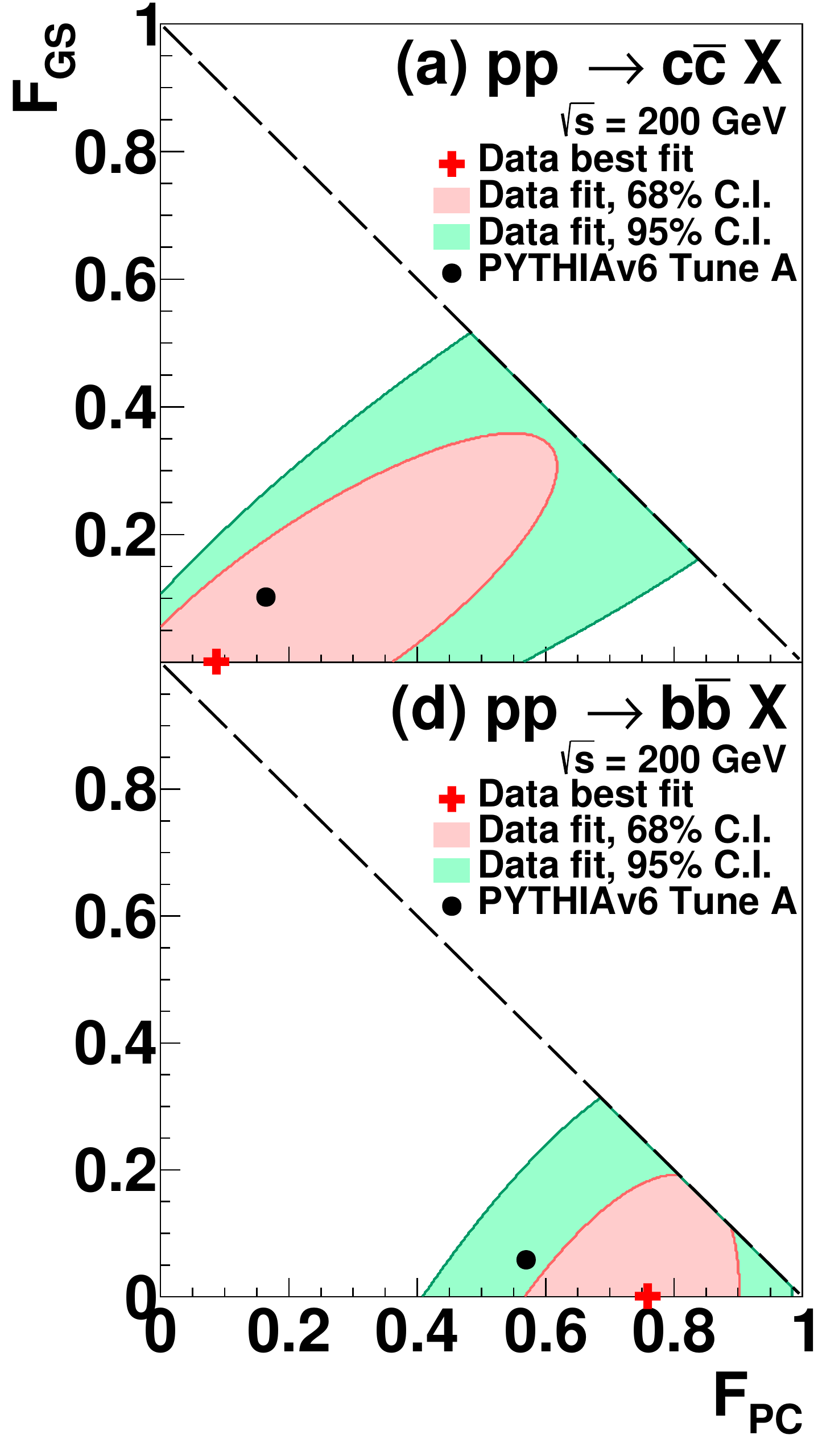}
\includegraphics[width=0.329\linewidth]{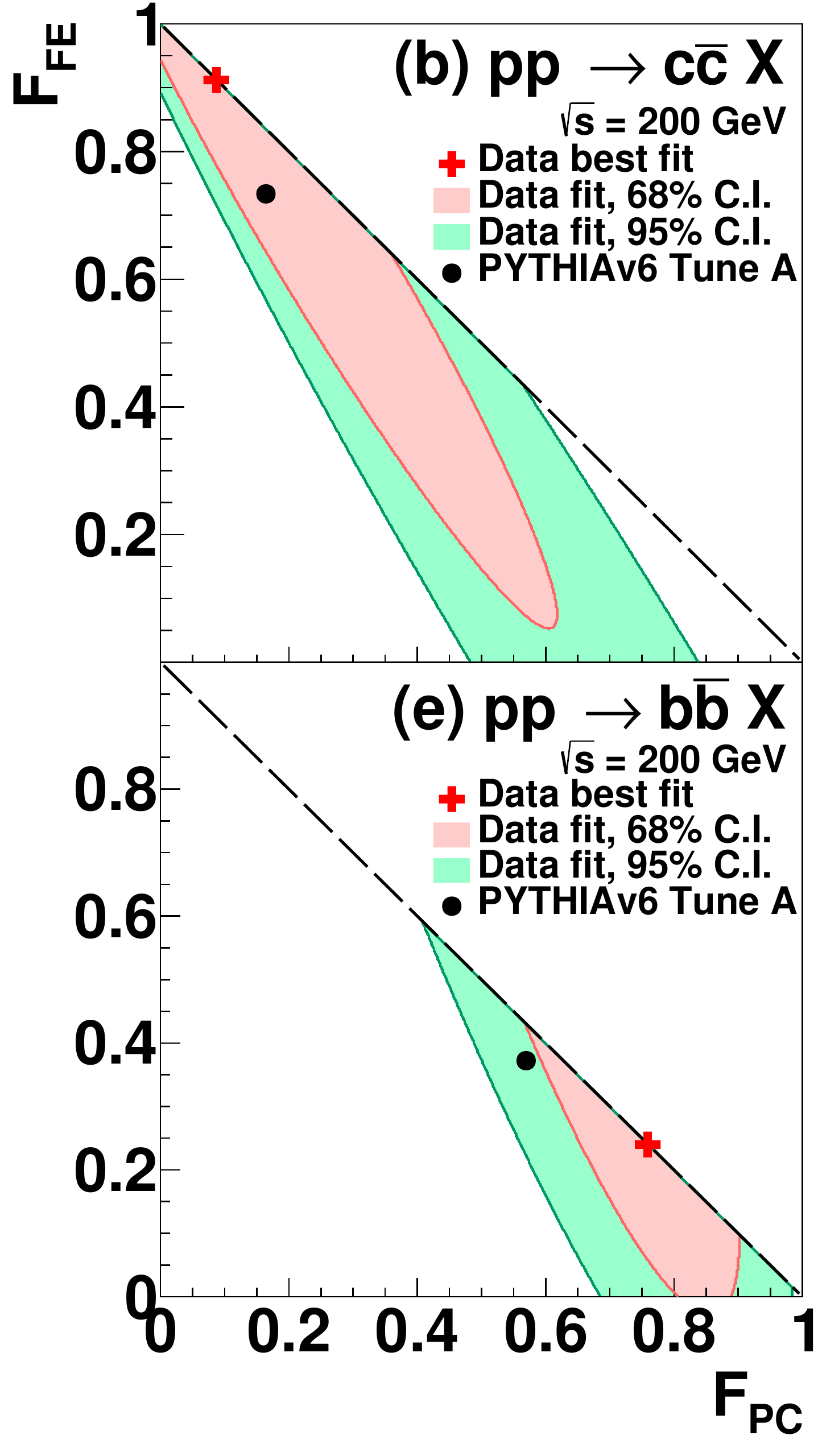}
\includegraphics[width=0.329\linewidth]{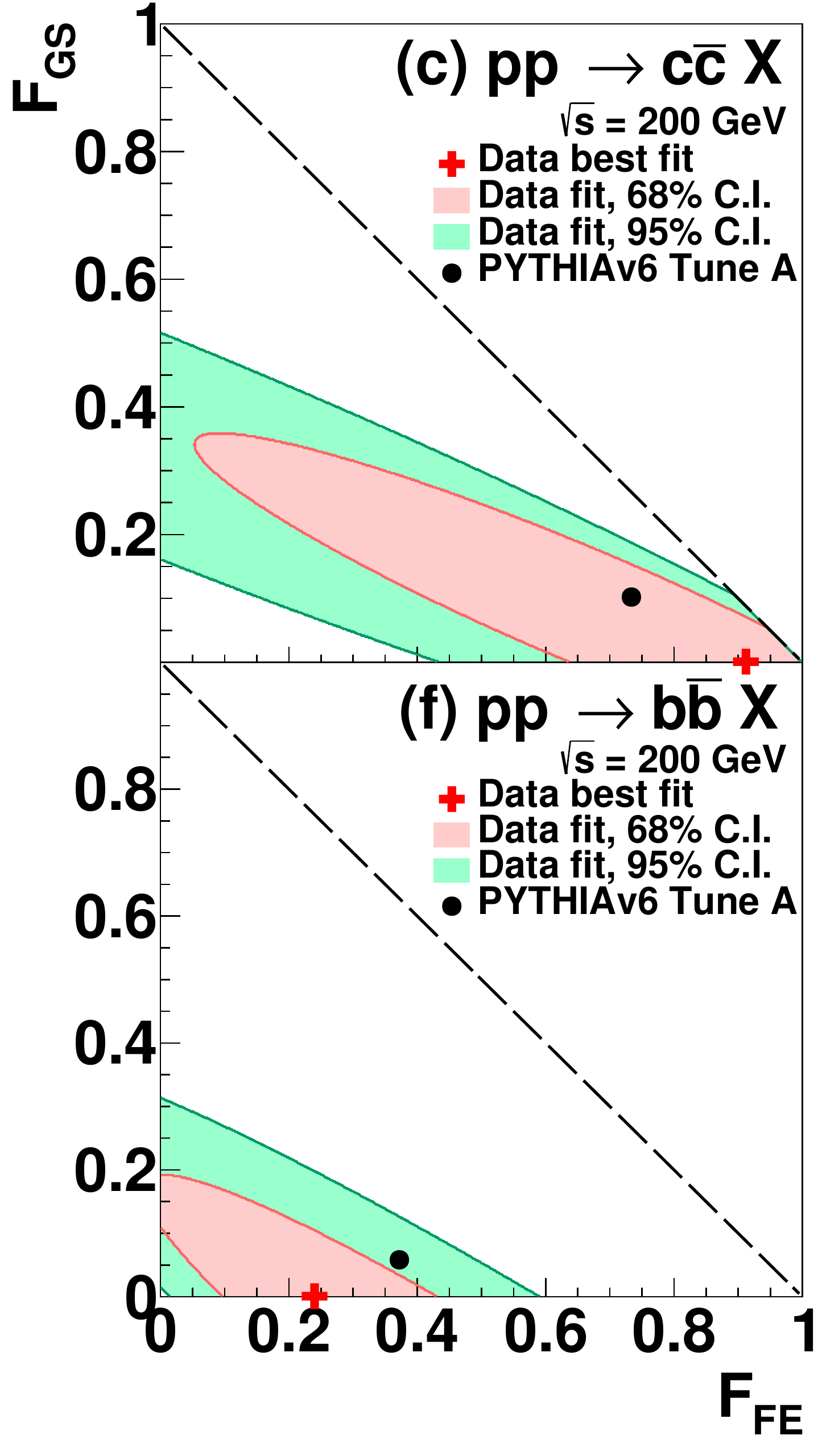}
\caption{\label{fig:ccfit} Credible intervals for (a,b,c) \cc and  
(d,e,f) \bb production mechanisms extracted from data and \pythia Tune A.}
\end{figure*}

To compute $P(\fraction|\D)$ from Eq.~\ref{eq:bayes1}, we adopt a Monte 
Carlo approach, in which multiple sets of nuisance parameters 
$\textbf{n*}$ are randomly generated. For each set of $\textbf{n*}$, the 
data \D are perturbed according to $\textbf{n*}$, and \sigmahf is 
constrained via a 1-parameter $\chi^{2}$ fit to the data. The posterior 
probability density $P(\fraction, \sigmahf, \textbf{n*}|\D)$ is then 
summed over different sets of $\textbf{n*}$ and normalized to unity in 
order to obtain $P(\fraction|\D)$. Finally, we construct $68\%$ and 
$95\%$ credible intervals from $P(\fraction|\D)$; boundaries of the 
intervals are contours of the posterior probability density 
$P(\fraction|\D)$.

The final results are presented in Fig.~\ref{fig:ccfit} in different 
projections of \fraction. For \cc (\bb), the \pythia Tune A 
implementation lies within the $68\%~(95\%)$ credible intervals obtained 
from our analysis. For the case of \cc, a positive correlation is 
observed between $F_{\rm PC}$ and $F_{\rm GS}$, both of which are individually 
anti-correlated with $F_{\rm FE}$. This is explained by the observation that 
the data sets can be reasonably well described by the following two 
cases: $\fraction=(0\%,100\%,0\%)$ and $\fraction=(62\%,0\%,38\%)$. From 
the posterior probability distributions, it is observed that the 
hierarchy $F_{\rm FE}>F_{\rm PC}>F_{\rm GS}$ is favored, consistent with the 
expectation from \pythia.

In contrast to \cc, PC is clearly the dominant $(76\%\pm^{14}_{19}\%)$ 
production process for \bb. Compared to \cc, the ordering of 
contributions from of PC and FE is reversed $F_{\rm PC}>F_{\rm FE}>F_{\rm GS}$, 
again consistent with the expectation from \pythia. The reversal in the 
hierarchy for \bb arises from the larger $b$ quark mass, which sets more 
demanding kinematic requirements for NLO processes.

The upper limits corresponding to the $95\%$ credible intervals for 
$F_{\rm GS}$ for \cc and \bb are $52\%$ and $31\%$ respectively. These 
limits take into consideration extreme cases in which only PC and GS 
contribute to the yield but FE does not. Priors with extra physical 
considerations may be incorporated to impose more stringent constraints 
in \fraction, however this is beyond the scope of our study.

In summary, we have presented an analysis of angular correlations of 
\mumu, \emu, and \ee pairs from \cc and \bb measured in \pp collisions 
at \s{200} at forward-forward, mid-forward, and mid-midrapidity, 
respectively. All measured angular correlations can be consistently 
described by distributions obtained from \pythia Tune A. In contrast, 
angular correlations generated using \powheg are broader than those from 
data.

Based on \pythia Tune A, we have performed a shape analysis using the 
combined data on heavy flavor angular correlations at \s{200}. This 
analysis constrains the relative contributions of the leading order pair 
creation, and next-to-leading order flavor excitation and gluon 
splitting processes, separately for \cc and \bb. The data indicate that 
the dominant production mechanism of \bb production is pair creation, 
and supports the scenario in which flavor excitation dominates \cc 
production. Similar measurements in $p$+$p$ collisions at different 
energies will provide insight on the energy dependence of heavy quark 
production mechanisms.

At RHIC energies, heavy quarks can be utilized to study initial gluon 
dynamics due to the small fraction of gluon splitting contribution. 
Besides $p$+$p$ collisions, heavy quarks are commonly used to study 
nuclear matter effects in $p$$+A$ and A+A collisions with the assumption 
that heavy quarks are mostly produced in the early stages of collisions. 
Similar measurements in $p$$+A$ may shed light on process dependent cold 
nuclear matter effects. A solid understanding of the contributions of 
heavy flavor processes in $p$+$p$ and $p$$+A$ collisions will be critical 
to precisely interpret results in A+A collisions, which suffer 
complications due to the contribution from gluon splitting process, 
particularly at LHC energies~\cite{Huang:2013vaa,Huang:2015mva}.


\begin{acknowledgments}

We thank the staff of the Collider-Accelerator and Physics
Departments at Brookhaven National Laboratory and the staff of
the other PHENIX participating institutions for their vital
contributions.  We acknowledge support from the 
Office of Nuclear Physics in the
Office of Science of the Department of Energy,
the National Science Foundation, 
Abilene Christian University Research Council, 
Research Foundation of SUNY, and
Dean of the College of Arts and Sciences, Vanderbilt University 
(U.S.A),
Ministry of Education, Culture, Sports, Science, and Technology
and the Japan Society for the Promotion of Science (Japan),
Conselho Nacional de Desenvolvimento Cient\'{\i}fico e
Tecnol{\'o}gico and Funda\c c{\~a}o de Amparo {\`a} Pesquisa do
Estado de S{\~a}o Paulo (Brazil),
Natural Science Foundation of China (People's Republic of China),
Croatian Science Foundation and
Ministry of Science and Education (Croatia),
Ministry of Education, Youth and Sports (Czech Republic),
Centre National de la Recherche Scientifique, Commissariat
{\`a} l'{\'E}nergie Atomique, and Institut National de Physique
Nucl{\'e}aire et de Physique des Particules (France),
Bundesministerium f\"ur Bildung und Forschung, Deutscher Akademischer 
Austausch Dienst, and Alexander von Humboldt Stiftung (Germany),
J. Bolyai Research Scholarship, EFOP, the New National Excellence
Program ({\'U}NKP), NKFIH, and OTKA (Hungary),
Department of Atomic Energy and Department of Science and Technology 
(India),
Israel Science Foundation (Israel), 
Basic Science Research Program through NRF of the Ministry of 
Education (Korea),
Physics Department, Lahore University of Management Sciences (Pakistan),
Ministry of Education and Science, Russian Academy of Sciences,
Federal Agency of Atomic Energy (Russia),
VR and Wallenberg Foundation (Sweden), 
the U.S. Civilian Research and Development Foundation for the
Independent States of the Former Soviet Union, 
the Hungarian American Enterprise Scholarship Fund,
the US-Hungarian Fulbright Foundation,
and the US-Israel Binational Science Foundation.

\end{acknowledgments}



%
 
\end{document}